\documentstyle[12pt,fleqn]{article}
\renewcommand{\baselinestretch}{1.2}

\newcommand{\bea}{\begin{eqnarray}}
\newcommand{\beq}{\begin{equation}}
\newcommand{\eea}{\end{eqnarray}}
\newcommand{\eeq}{\end{equation}}
\newcommand{\nnu}{\nonumber}
\newcommand{\di}{\mbox{d}}
\newcommand{\half}{\frac{1}{2}}
\newcommand{\spav}[1]{\parbox{1mm}{\vspace*{#1}}}

\begin{document}

\begin{titlepage}
\begin{flushright}
SISSA/ISAS 128--92--EP\\
July 1992
\end{flushright}
\spav{1cm}\\
\begin{center}
{\LARGE\bf \sc The Gravitational Field of  String\\}
{\LARGE\bf \sc Matter when the Dilaton is Massive \\}
\spav{1cm}\\
{\large  Marco Fabbrichesi and Roberto Iengo}
\spav{2cm}\\
{\em International School for Advanced Studies
(SISSA/ISAS)}\\
and \\
{\em INFN, Sezione di Trieste}
\spav{.5cm}\\
{\em via Beirut, 2-4, I-34014 Trieste, Italy.} \\
\spav{2cm}\\
{\sc Abstract}
\end{center}
We study numerically the gravitational field of a star
made of massive and neutral string states for the case in
which the dilaton is massive. The solution exhibits very
simple scaling properties in the dilaton mass. There is no
horizon and the singularity is surrounded by a halo
(the physical size of which is
inversely proportional to the dilaton mass)
where the scalar curvature is very large and proportional to the square
of the dilaton mass.

\vfill
\end{titlepage}

\newpage
\setcounter{footnote}{0}
\setcounter{page}{1}

{\bf 1.} Neutral and scalar string excitations which are
massive already in ten dimensions~\cite{GSW} represent the simplest
form of that exotic state of matter in which the constituents are string
states not belonging to  the massless sector of the
string spectrum to which instead
 ordinary matter, like  quarks and leptons, presumably belongs (their
 masses originating through some
still unknown symmetry breaking mechanism).

The physical relevance of such string matter stems
 from several
studies~\cite{lavori} which indicate that, at high energy density, the
most probable configuration in string theory is the one in which most of
the massive states are
excited. According to this view, an electrically neutral
collapsing star of sufficiently large mass
 would start out as a celestial body made of
ordinary matter but would
eventually evolve into a {\em string star} in which most of its mass is
now carried by neutral string excitations.

It is therefore of some interest, we believe, to study the gravitational field
around such
a string star.

The scalar, neutral and massive string excitations
couple only to gravitons and dilatons; therefore, the gravitational
field around a star made of a large number of such states
 is
described by Einstein's equations in which the energy-momentum tensor is
the one for a
scalar field. In the original string scenario,
these equations arise as conditions for the vanishing of
the one-loop beta functional required
by Weyl invariance; the dilaton is massless
and the field equations admit an exact
solution~\cite{BD,new}. This solution, and its relationship to
string theory, has been discussed recently
in~\cite{FIR}, to which this letter is closely related. The most significant
feature of such a solution is the absence of horizon. The relevant elements
$g_{rr}$ and $g_{00}$ of the
space-time metric are shown in fig.1, where they are compared to the
Schwarz\-schild's ones. As it is possible to see from the figure, while the two
solutions are equivalent at large distances, they
are remarkably different closer to the gravitational
 radius: at the horizon
Schwarz\-schild's
$g_{rr}$ diverges  and $-g_{00}$, which is just its inverse,
crosses into negative values, whereas
the ones corresponding to the string star solution
are never either negative or infinite. These results seem to be in agreement
with previous work~\cite{Price} in which the existence of a static and
uncharged
static solution with a scalar field has been shown to be incompatible
with the presence of a non-singular horizon
(see, also, \cite{new}).

It is tempting to speculate that also in a realistic scenario,
derived from string theory, in which
ordinary matter, after obtaining a mass  much lower than Planck mass
by some yet unknown mechanism,
  continues to behave with respect to the dilaton
as it were in the massless sector of the string spectrum. According
to this view, the dilaton
field  remains massless as well as decoupled to ordinary matter
at large distance. Such a
possibility can be entertained without violating any
of the present experimental
observations. In this case all the results
of~\cite{FIR} would hold true.

In this letter we address the question of what happens to the solution
with a massless dilaton
in the opposite case in which the dilaton does couple to ordinary matter
and therefore---because of
the current astronomical bounds~\cite{bounds} on Newton's
inverse-square law---it  has to
become massive. Such an investigation is, for its
 very nature only
preliminary, because of our present ignorance of
 how the basic ingredients of the superstring, like world-sheet
conformal invariance, determine the structure of the effective low-energy
theory with a massive dilaton
(see~\cite{T} for a related discussion in cosmology).

The reader, confronted by a growing literature on the subject of
solutions to graviton plus dilaton (and, plus Maxwell fields) gravity
in four space-time dimensions,
should bear in mind that the general solution for a source with an
arbitrary coupling to
massless dilaton, graviton and Maxwell fields is contained
and discussed in the  papers of ref.~\cite{papers}.
Paper~\cite{horo} discusses the special case in which
the coupling to the massless dilaton
field is a function of the Maxwell charge and it vanishes  as the charge
goes to zero. Our solution is the one in which
the source is neutral with respect to the the Maxwell field
but nonetheless has a non-vanishing coupling to  the dilaton field.

\vspace{1cm}
{\bf 2.} The field equations are easily written as
variations of the action
\beq
 S =\frac{1}{2\kappa^2}\int \di ^4 x \sqrt{-g}\,
\left[ {\cal R}
  -
g^{\alpha\beta}\partial_\alpha \phi \partial_\beta \phi -
m^2 W( \phi ) \right] \, ,\label{action}
\eeq
and read
\bea
-e^{-\lambda}\left( \frac{\lambda '}{r} - \frac{1}{r^2} \right)
-\frac{1}{r^2} & = & - \frac{e^{-\lambda}}{2} (\phi ')^2 -
 \frac{m^2}{2} W(\phi)   \nnu \\
 e^{-\lambda}\left( \frac{\nu '}{r} + \frac{1}{r^2} \right)
-\frac{1}{r^2} & = &  \frac{e^{-\lambda}}{2} (\phi ')^2 -
 \frac{m^2}{2} W(\phi)  \nnu \\
\phi '' + \left( \frac{2}{r} + \frac{\nu ' - \lambda '}{2} \right)
\phi ' & = &  \frac{e^{\lambda}}{2} m^2 \frac{\delta W}{\delta \phi}
 \, .\label{eq1}
\eea
We have restricted ourselves to the static and spherically symmetric
case and therefore used the parametrization
\beq
g_{00} = -e^{\nu (r)} \qquad g_{rr} = e^{\lambda (r)}
\eeq
for the space-time metric,
the signature of which is $-\, +\, +\, +$;
 prime means differentiation with respect to
the radial coordinates, $\phi \equiv \kappa \varphi$
 is the rescaled dilaton field $\varphi$ and $\kappa^2 = 8\pi
G_N$, $G_N$ being Newton's constant.

In general,
\beq
W(\phi) = \phi ^2 + \alpha \phi ^3 +
\beta \phi ^4 + \ldots \label{pot}
\, ,
\eeq
where the first term gives rise to the mass, and the remaining ones to an
effective potential for the dilaton field.

Two questions arise.

The action~(\ref{action}) without the mass term  is
obtained from the usual one
\beq
S= \frac{1}{2\kappa^2} \int
\di x \sqrt{-g} e^{-2\phi} \left[ {\cal R}
+4 \nabla \phi  \cdot \nabla \phi \right]
\eeq
 by the rescaling
$g \rightarrow e^{2\phi}g$ and
 $ \phi \rightarrow \sqrt{2}
\phi /2$, and it represents
the low-energy effective theory one
 derives from string theory.
After symmetry breaking, other terms beside the mass $m^2\phi^2$
may appear and give rise to a potential  for the dilaton
field as in~(\ref{pot}). Consider, for instance,
a term like $\beta\phi^4$.
Even though the coupling to $\varphi ^4$ is typically of the order of
$G_Nm^2$ (that is, $\beta$ of order of one),
the net contribution to the equations of motion
is of the same order as the ones included in~(\ref{action}), as
one can readily see by inspection.

We have explicitly checked our solution against such terms
in the potential and verified that its nature,
qualitative features as well as most of
the numerical values are left unchanged for
values of the parameters
$\alpha$, $\beta$ up to order of ten. The only case in
 which there is a substantial change in behavior
  is  when we take, for example,
$\alpha > 0$ and $\beta$ and the coefficients of the
other even power terms vanishing, thus forcing the potential
 to become unstable (recall that $\phi$ is negative). Therefore,
 except for this
somewhat pathological case, we are entitled to
disregard for the sake of simplicity these
extra terms, and  take in
 what follows
$W(\phi) = \phi^2$ only.

A second question concerns possible string corrections to~(\ref{action}).
These are, for example, of the type $\alpha ' {\cal R}^2$, where $\alpha
'$ is
the inverse of the string tension. However, in this case, one finds
that even for very large values of the scalar curvature (as we are
in fact going to have) such terms are suppressed by a factor $\alpha '
m^2$ with respect to the term linear in the curvature ${\cal R}$, and can
accordingly be neglected altogether.

\vspace{1cm}
{\bf 3.}  Equations~(\ref{eq1}) do not seem amenable of any
exact solution. We
study them by numerical methods.

It is convenient to recast~(\ref{eq1}) as a system of four first-order
differential equations:
\bea
y_1 ' & = & \frac{1}{\rho} \frac{y_3}{y_4} \nnu \\
y_2 ' & = & \frac{1}{2\rho} \frac{y_2 y_3^2}{y_4^2} \nnu \\
y_3 ' & = & \xi^2 \rho^2  y_1 y_2\nnu \\
y_4 ' & = & y_2 \left( 1 - \half \xi^2 \rho^2 y_1^2
\right)  \label{eq2}
\eea
by means of four dimensionless functions:
\bea
y_1 & = & \phi \nnu \\
y_2 & = & e^{(\nu+\lambda)/2} \nnu \\
y_3 & = & \rho e^{(\nu-\lambda)/2} \nnu \\
y_4 & = &  \rho^2 e^{(\nu-\lambda)/2}
\phi '\, .
\eea
In ~(\ref{eq2}), $\rho = r/r_g$, where $r_g=2G_NM$ is the gravitational
radius of the star. The prime is now differentiation with respect
to the coordinate $\rho$.

The parameter $\xi = mr_g$ is
very large for the physically interesting scenario in which the
Compton wavelength of the dilaton $1/m$ is much shorter than
the gravitational radius; we have representatively considered values of $\xi$
between $10$ to $10^{10}$.

We set the boundary conditions at large distance
for (\ref{eq2}) as follows:
\bea
y_2 & \rightarrow & 1 \nnu \\
y_4 & \rightarrow & \rho -1 \, ,
\eea
that is, we require that both $g_{00}$ and $g_{rr}$
must go into Schwarz\-schild metric for sufficiently large distances.

Less clear is the asymptotic value for the dilaton field $\phi$.
The exponential decay
\beq
y_1 \rightarrow \phi_0\, e^{-\xi\rho}/\rho \, ,\label{exp}
\eeq
that we expect by solving the corresponding equation,
leaves the constant $\phi_0$ undetermined. While for the massless
case this indeterminacy can be resolved
by comparison with string perturbation
theory~\cite{FIR}, the problem seems more complicated here.
Luckily, as we shall see, the solution we find is very stable
over a large range of changes in the initial condition for the dilaton
field. We therefore fix it
 by assuming in the region $\rho -1 \ll 1$ that
\beq
y_1 \approx K_0 \left( 2\xi \sqrt{\rho -1} \right) \, ,
\eeq
as it is suggested by the study of the dilaton equation in a fixed
background, and putting by hand the numerical
coefficient in front of the
Bessel function. This choice makes the dilaton
field behave as in~(\ref{exp}) for $\rho -1 \gg 1$, and as
\beq
y_1 \approx \frac{\exp -2\xi \sqrt{\rho -1}}{2\xi \sqrt{\rho -1}}
\eeq
and
\beq
y_1 \approx \log \left( 2\xi \sqrt{\rho -1} \right)
\eeq
for, respectively, $\xi^{-2} \ll \rho -1 \ll 1$ and
$0< \rho -1 \ll \xi^{-2}$.

This procedure amounts to placing the position at which the dilaton field
begins to be different from zero. Within an acceptable range of
initial values, once it
has been fixed, the dilaton
field rapidly goes into its asymptotic (and universal) behavior
beyond the gravitational
radius.

\vspace{1cm}
{\bf 4.} We can thus solve the system of first order differential
equations~(\ref{eq2}) as an initial-value problem
 by means of a standard program~\cite{NAG} based on a variable-order,
 variable-step method implementing the Backward
Differential Formulas.

The results of the numerical integration
 are displayed in
figs. 2a-2b for the metric and 3 for the dilaton field, where they
are plotted  for the special value $\xi = 10$.
 Although such a
small value is not physically relevant, it is useful because it makes the
plotting more readable. Moreover, the solution exhibits very simple
scaling properties that allow to infer from the plot of figs. 2a-2b and 3 the
corresponding ones for higher, and more realistic, values of $\xi$.

As we have already pointed out,
we obtain a solution which is surprisingly stable with respect to
variations in
the dilaton mass, the asymptotic value of $\phi$ and the parameters in
$W(\phi)$. The features of the solution we would like to
emphasize are that:
\begin{itemize}
\item
like in the massless case, the horizon is absent;
\item
there is a very narrow region near the would-be horizon where there is a
very sharp transition away from Schwarz\-schild's solution;
\item
outside this region---that is, for $r>r_g$---the field $\phi$ can be
neglected and the solution is indistinguishable from the Schwarz\-schild's
one;
\item
inside this region---that is, for $r<r_g$---the solution becomes stable and
depends very little on details.
\end{itemize}

By varying the input value of $\xi$ and comparing the outputs
so obtained we have
verified that
the behavior of the space-time metric
for $\rho \rightarrow 0$ is the following:
\bea
g_{rr}  & \simeq & \frac{1}{\xi^{2}} P_0 \rho ^{Q_0^{2} /2 + 1} \nnu \\
-g_{00}  &  \simeq & \frac{1}{\xi} S_0 \rho ^{Q_0^{2} /2 - 1} \,
 \label{a}
\eea
 while the dilaton field remains unchanged and:
 \beq
 \phi \simeq Q_0 \log \rho - C_0 \, . \label{b}
 \eeq

Also the scalar curvature  behaves in a simple manner:
\beq
{\cal R} \simeq \xi^2  \left\{ \frac{Q_0^2}{P_0} \rho^{-(3 + Q_0^2 /2)}
 + O(\log \rho)
\right\} \, .\label{c}
\eeq

The numerical values for $Q_0$, $C_0$ and $P_0$ seem
to be universal,
independent of the dilaton mass and every other parameter; they turn out to be
$Q_0 \simeq 2.5$, $C_0 \simeq -1$ and
$P_0 \simeq 4$.
 $S_0$ seems to depend more on the
input conditions, however being always
of order one. Notice that the
solution for the massless case~\cite{FIR} was of
the same kind with $Q_0^2 =
2(\sqrt{2}+1)/(\sqrt{2}-1)$.

The behaviors~(\ref{a})-(\ref{c}) are well satisfied inside the Schwarz\-schild
radius, whereas away from it the solution is more sensitive
to the initial conditions and does not follow a simple scaling law in $\xi$.

\vspace{1cm}
{\bf 5.} Let us briefly discuss the solution
we have found. It resembles in a qualitative
manner the massless case. Away from the would-be horizon, the metric is
indistinguishable from Schwarz\-schild's. However, when distances of the
order of $1/\xi^2$ are reached, the solution departs dramatically from
Schwarz\-schild's and the horizon disappears.  For large values of $\xi$,
this change takes place in a very thin shell around $r_g$. This
can be compared with the
massless case in which the departure occurs at distances of the order of
$r_g$ instead.

 At the same
time, the scalar curvature (see fig.4) grows in a very steep manner and
produces a high-curvature sphere around the  singularity.
This is to be expected since in this region $\phi$ changes from
being vanishingly small to being
of order one, thus acting back on the space-time
metric
and giving a large contribution to the curvature as well.
Fig.4 makes clear how the massive case differs from the massless one
only in that the growth of the curvature is more abrupt (again, it takes
place in the region of size $1/\xi$ before $r_g$) and steeper. They
become equivalent closer to the singularity where the dilaton field
becomes effectively massless in any case.

 The
physical size of the halo of  large curvature  is
\beq
\int_0^1 \di \rho \sqrt{g_{rr}} \simeq 1/\xi
\eeq
and therefore of order $1/m$ in standard units.

We feel that these results are very natural. The presence of a mass
for the dilaton field, by making available a second parameter to the problem,
changes the scale at which the departure from Schwarz\-schild solution
takes place and also the magnitude of such departure;
it does not  however change the nature
of it, that remains the one found in the massless case.

\newpage
\renewcommand{\baselinestretch}{1}

\newpage
{\Large \bf Captions}\\
\vspace*{1cm}\\
Fig.1: Schwarzschild's space-time metric compared to the string star's for a
massless dilaton field.\\
\vspace*{1cm}\\
Fig.2a: Schwarzschild's $g_{00}$ compared to the string star's for a massive
dilaton field.\\
\vspace*{1cm}\\
Fig.2b: Schwarzschild's $g_{rr}$ compared to the string star's for a massive
dilaton field.\\
\vspace*{1cm}\\
Fig.3: The massive dilaton field.\\
\vspace*{1cm}\\
Fig.4: The scalar curvature: Massless vs. massive dilaton field.


\begin{thebibliography}{99}
{\small

\bibitem{GSW}  See, for example, M. Green, J. Schwarz and E. Witten, {\em
Superstring
Theory} (Cambridge University Press, Cambridge 1987).

\bibitem{lavori} S. Frautschi, {\em Phys. Rev} {\bf D3} (1971) 281;\\
R.D. Carlitz, {\em Phys. Rev.} {\bf D5} (1972) 3231;\\
D. Mitchell and N. Turok, {\em Nucl. Phys.} {\bf B294} (1987) 1138;\\
M.J. Bowick and B. Giddings {\em Nucl. Phys.} {\bf B325}
 (1989) 631;\\
F. Lizzi and A. Sanda, {\em Nucl. Phys.} {\bf B359} (1991) 441.

\bibitem{BD} C. Brans, {\em Ph.D. Thesis}, Princeton University, New
Jersey 1961 (unpublished);\\
 C. Brans and R.H. Dicke, {\em Phys. Rev.} {\bf 124} (1961)
925;\\
C. Brans, {\em Phys. Rev.} {\bf 125} (1962) 2194.

\bibitem{new}
A.I. Janis, E.T. Newman and J. Winicour, {\em Phys. Rev. Lett.}
{\bf 20} (1968) 878.

\bibitem{FIR} M. Fabbrichesi, R. Iengo and K. Roland,
{\em The Gravitational Field of String Matter}, preprint
SISSA/ISAS 52-92-EP;\\
see, also: M. Fabbrichesi and R. Iengo, {\em Phys. Lett} {\bf B264} (1991)
319.

\bibitem{Price} W. Israel, {\em Phys. Rev} {\bf 164} (1967) 1776;\\
R.H. Price, {\em Phys. Rev.} {\bf D5} (1972) 2419.

\bibitem{bounds} A. De Rujula, {\em Nature} {\bf 323} (1986) 760.

\bibitem{papers} P. Dobiasch and D. Maison, {\em Gen. Rel. and Grav.}
{\bf 14} (1982) 231;\\
A. Chodos and S. Detweiler, {\em Gen. Rel. and Grav.} {\bf 14}
(1982) 879;\\
G.W. Gibbons, {\em Nucl. Phys.} {\bf B207} (1982) 337;\\
G.W. Gibbons and M.J. Perry, {\em Nucl. Phys.} {\bf B248} (1984) 629;\\
G.W. Gibbons and K. Maeda, {\em Nucl. Phys.} {\bf B298} (1988) 741.

\bibitem{horo} D. Garfinkle, G.T. Horowitz and
A. Strominger, {\em Phys. Rev} {\bf D43} (1991) 3140.

\bibitem{T} A.A. Tseytlin, {\em String Cosmology and Dilaton},
to appear in {\em Int. J. Mod. Phys.} {\bf A} (1992).

\bibitem{NAG} D02EBF, NAG Fortran Library--Mark 15
(NAG Ltd.,
Oxford, U.K., 1991).


    }

\end{thebibliography}
\end{document}